\documentclass{article}

\usepackage{PRIMEarxiv}

\usepackage[utf8]{inputenc} 
\usepackage[T1]{fontenc}    
\usepackage{hyperref}       
\usepackage{url}            
\usepackage{booktabs}       
\usepackage{amsfonts}       
\usepackage{nicefrac}       
\usepackage{microtype}      
\usepackage{lipsum}
\usepackage{fancyhdr}       
\usepackage{graphicx}       
\usepackage{amsmath}
\usepackage{algorithm}
\usepackage{algorithmic}
\usepackage{multirow} 
\graphicspath{{media/}}     

\title{Contrastive-KAN: A Semi-Supervised Intrusion Detection Framework for Cybersecurity with scarce Labeled Data
}

\author{
  Mohammad Alikhani \\
  Faculty of Electrical Engineering \\
  K.N. Toosi University of Technology \\
  Tehran, Iran\\
  \texttt{m.alikhani2@email.kntu.ac.ir} \\
   \And
  Reza Kazemi \\
  Faculty of Electrical Engineering \\
  K.N. Toosi University of Technology \\
  Tehran, Iran\\
  \texttt{Rezakazemi@email.kntu.ac.ir} \\
}

\begin{document}
\maketitle

\begin{abstract}
In the era of the Fourth Industrial Revolution, cybersecurity and intrusion detection systems are vital for the secure and reliable operation of IoT and IIoT environments. A key challenge in this domain is the scarcity of labeled cyberattack data, as most industrial systems operate under normal conditions. This data imbalance, combined with the high cost of annotation, hinders the effective training of machine learning models. Moreover, the rapid detection of attacks is essential, especially in critical infrastructure, to prevent large-scale disruptions. To address these challenges, we propose a real-time intrusion detection system based on a semi-supervised contrastive learning framework using the Kolmogorov-Arnold Network (KAN). Our method leverages abundant unlabeled data to effectively distinguish between normal and attack behaviors. We validate our approach on three benchmark datasets, UNSW-NB15, BoT-IoT, and Gas Pipeline, using only 2.20\%, 1.28\%, and 8\% of labeled samples, respectively, to simulate real-world conditions. Experimental results show that our method outperforms existing contrastive learning-based approaches. We further compare KAN with a traditional multilayer perceptron (MLP), demonstrating KAN's superior performance in both detection accuracy and robustness under limited supervision. KAN's ability to model complex relationships, along with its learnable activation functions, is also explored and visualized, offering interpretability and the potential for rule extraction. The method supports multi-class classification and proves effective in safety, critical environments where reliability is paramount.
\end{abstract}


\section{Introduction}\label{sec_Introduction}
Today, with the rapid development of networking, the Internet of Things (IoT) and the Industrial Internet of Things (IIoT), including technologies such as smart home applications, medical devices, automated vehicles, and sensor networks \cite{abed2024modified}, are being increasingly adopted. As a result, cybersecurity has become one of the most critical aspects of both cyberspace and network infrastructures. The need for secure digital environments is evident in the era of the Fourth Industrial Revolution, where internet connectivity is integrated into many aspects of modern life \cite{muneer2024critical}, \cite{mukherjee1994network}, \cite{yick2008wireless}.

On the other hand, in the era of machine learning (ML) and deep learning (DL) has found its way in many other domains as in we can find them in the field of navigation \cite{alikhani2025long}, image processing \cite{lu2007survey}, medical field such as personalized treatment \cite{wen2023deep}, intelligent applications in finance \cite{gadre2016review}, industrial automation \cite{iqbal2019fault} and many other applications .

In IoT and IIoT scenarios, intrusion detection systems are designed to identify suspicious patterns in sensor, actuator, and network packet data. With the advent of , the field of ML and DL intrusion detection has shifted from classical rule-based methods toward intelligent data-driven approaches. ML techniques require fewer computational resources and less data compared to DL methods. However, they often lack the generalizability needed to handle unseen data. In contrast, DL methods offer superior performance but require large volumes of data and significantly more computational resources \cite{muneer2024critical}.

To address these challenges, researchers have proposed various solutions. Unsupervised methods aim to reduce the cost and time associated with data annotation by leveraging unlabeled data \cite{li2024genos}. However, these approaches usually fall short in achieving the desired performance. To preserve data privacy, federated learning techniques have been introduced to train models without the need to centralize data \cite{muneer2024critical}. Despite these advances, many existing methods still lack the interpretability required by industry for safe and reliable operation \cite{adadi2018peeking}.

To address challenges such as data augmentation, labeled data scarcity, and data imbalance, many algorithms rely on oversampling techniques such as the Synthetic Minority Over-sampling Technique (SMOTE) \cite{lachure2024securing} and its variants. Generative methods using Generative Adversarial Networks (GANs) have also been explored. However, there are several issues that limit the effectiveness of such techniques. First, these methods often experience performance degradation when the imbalance ratio is large, which is a common characteristic of intrusion detection datasets. Second, in many of these datasets, the data distributions of attack and normal samples are quite similar. As a result, oversampling techniques can lead to blurred boundaries between attack and normal classes. In the case of GANs, training is often unstable and difficult to converge. In this work, we discourage the use of these methods for handling data imbalance and augmentation. Instead, we adopt a masking strategy that randomly masks out a certain percentage of the data.

Another commonly used technique in intrusion detection is feature selection \cite{kravchik2021efficient}. However, a significant limitation of this approach is that if an attack targets a specific feature that has been removed during the selection process, the model may fail to detect it. In this work, we avoid using feature selection to preserve potentially important information. The only exception is the removal of features with no variation in the Gas Pipeline dataset. No other form of feature selection is applied.

One of the major challenges in real-time intrusion detection is preprocessing, which can delay the timely detection of cyber intrusions. Many methods use window-based sampling approaches. The problem with such techniques is that the algorithm must wait for a complete data window before proceeding with operations such as averaging or applying one-dimensional convolutional neural networks (1D-CNNs) for feature extraction and final classification. Furthermore, window-based algorithms often exhibit lower performance \cite{de2020intrusion}. Another commonly used preprocessing method is principal component analysis (PCA), which reduces data dimensionality by transforming the input into a set of orthogonal components that capture maximum variance along the principal components. Our approach, in contrast, processes individual data samples, eliminating the need for windowing. The associated preprocessing consists only of feature encoding and standard scaling for normalization. This preserves the interpretability of features and enables timely intrusion detection.

As discussed earlier, the limited availability of annotated data is one of the biggest challenges in machine learning (ML) and deep learning (DL). We have outlined the advantages and disadvantages of both ML and DL. To bridge the gap between these approaches and to address the issue of limited labeled data, semi-supervised methods have been proposed. In these methods, the network is first pretrained on large amounts of unlabeled data, and then fine-tuned using only a small labeled subset. One example of such semi-supervised learning is the SimCLR framework introduced in \cite{chen2020simple}, which we adopt in this work using the Kolmogorov-Arnold Network (KAN). In our study, a feature extractor based on a single layer of KAN is trained using abundant unlabeled data. These data points are first augmented into two different views, and the feature extractor is trained to produce similar representations for these augmented views, which are essentially different versions of the same data point. Once pretraining is completed, fine-tuning is performed using only a small labeled subset.

To address the interpretability and explainability challenges faced by deep learning models, we incorporate the Kolmogorov-Arnold Network (KAN) into our framework. First proposed in \cite{liu2024kan}, KAN is a variant of a fully connected layer where traditional weights are replaced with learnable activation functions represented as splines. These splines approximate univariate functions capable of capturing complex patterns in the data. As will be discussed in later sections, KAN outperforms multilayer perceptron (MLP) networks with the same number of parameters. KAN decomposes complex multivariate inputs into multiple univariate functions. Originally, it was used for solving ordinary and partial differential equations \cite{yeo2024kan}, but recent studies have extended KAN's applications to time series analysis \cite{liu2025novel} and classification tasks.

The contributions of this study are as follows:
\begin{itemize}
    \item An efficient real-time intrusion detection system based on semi-supervised contrastive learning, which leverages large amounts of unlabeled data.
    \item Utilizing only a small fraction of labeled data while achieving better performance than other state-of-the-art methods.
    \item Investigating the interpretability of the proposed contrastive-KAN method through visualization of the learned spline activation functions.
    \item Conducting a thorough ablation study and comparison with traditional MLP networks, exploring different hyperparameters of the KAN model to identify those that most impact classification results.
\end{itemize}

The remainder of the paper is organized as follows: \autoref{sec:Related Works} reviews related work in the literature; \autoref{sec:Datasets} provides a detailed description of the datasets, including key characteristics, features removed, and data ratios; \autoref{sec:Method} outlines the problem formulation, the Kolmogorov-Arnold representation theory and KAN, as well as the contrastive training and fine-tuning stages. \autoref{sec_Results} presents the experimental results, followed by an extensive ablation study investigating KAN's hyperparameters, including spline order, grid size, and network width. Visualization of the activation functions learned by the pre-trained feature extractor and a comparative analysis with other methods are also included. Finally, \autoref{sec_Conclusion} concludes the paper and summarizes the key findings.

\section{Related Works}\label{sec:Related Works}
In \cite{wang2022cyber}, different ensemble learning approaches, such as XGBoost, random forest, support vector machine (SVM), and stacked deep learning (DL) models, are evaluated for cyber-attack detection. This study highlights the effectiveness of tree-based models, such as XGBoost, in comparison to DL models. The results indicate that stacked DL models exhibit greater sensitivity to outliers compared to ensemble methods. In \cite{wang2022stacked}, a stacked DL model is also utilized for detecting cyber-attacks in SCADA systems by combining the outputs of five individual DL models. This approach achieves higher performance compared to individual models. Moreover, this method employs random forest to determine feature importance.

In \cite{zhang2025network}, a CNN-LSTM hybrid network is proposed to improve temporal dependency modeling in IIoT networks. However, the authors acknowledge that challenges related to data imbalance and privacy remain unresolved, and they suggest adopting federated learning as a direction for future research. A probabilistic hybrid ensemble classifier (PHEC) based on federated learning for IoT network intrusion detection is proposed in \cite{chatterjee2022federated}. While this framework achieves high detection rates, it appears that, for the multi-class Gas Pipeline dataset, only a subset of classes was considered rather than all classes, limiting fair comparison with other state-of-the-art methods. Furthermore, the use of order reduction operations in this work may introduce additional computational complexity. In \cite{li2021sustainable}, the authors introduce a sustainable ensemble learning framework for intrusion detection in industrial control systems (ICS), which utilizes sensitivity to different attack types and historical data reuse. In this framework, models are assigned dynamic importance scores based on their effectiveness in detecting different attack types.

A bidirectional simple recurrent unit (BiSRU) model is introduced in \cite{ling2021intrusion} to address the limitations of traditional recurrent neural networks (RNNs), particularly the vanishing gradient problem, by using skip connections and a bidirectional structure. This BiSRU model improves training efficiency and enhances the model's ability to capture both forward and backward temporal dependencies. Although this model demonstrates effectiveness and efficiency, its performance is highly sensitive to data balance.

\cite{huda2018securing} proposes an ensemble-based detection model combining deep belief networks (DBNs) with support vector machines (SVMs) to address cyber-security challenges by analyzing network traffic and payload features. One advantage of this method is the one-class training of the SVM model, which can be performed using only normal data. However, DBNs still require a large amount of labeled data. Additionally, the scalability and training time of SVM for large datasets pose significant challenges for the adoption of this framework in real-world applications.

The work in \cite{Jia2024SWaTPSaEP} focuses on reducing false positive alarms by utilizing a graph-guided masked autoencoder (GGMAE) for anomaly detection in the process industry. This approach employs a topology graph to simulate process models and applies the Kullback-Leibler divergence as a loss function to ensure input-output consistency.

In \cite{xia2025condgad}, contrastive learning with a multi-augmentation strategy is employed to encode dynamic graph embeddings. These embeddings are subsequently passed through a transformer-based module to capture evolving patterns. Then, a forecasting model predicts long-term trends for anomaly detection. \cite{chen4757427harnessing} utilizes a feature encoder that converts input sequences into hidden representations, and a neural transformation module that generates multiple latent views. Training is performed using two contrastive losses: a contextual neural contrastive loss that aligns contextual information, and a discriminative contrastive loss that ensures diversity while preserving semantic consistency.

In \cite{Guo2025SWaTPSaEP}, the authors propose an unsupervised hierarchical strategy-based global-local graph contrastive learning (HS-GLGCL) framework to address the challenge of limited labeled data. This approach first constructs a process topology graph based on prior knowledge, then enhances it through graph augmentation and applies graph contrastive learning to extract meaningful representations from the unlabeled data. Finally, anomalies are identified based on reconstruction error thresholds and Kullback-Leibler divergence. \cite{belay2024self} introduces an encoder for reinterpretation learning, a decoder for latent reconstruction, a memory network for pattern recognition, and a denoiser for input correction, which is pre-trained by self-supervised learning. While the approach is comprehensive, the complexity of the framework can hinder real-time intrusion detection.

\cite{choi2024self} proposes a time-series anomaly detection framework based on learnable data augmentation and contrastive learning in a self-supervised setting. The main innovation of this framework is the learnable data augmentation module that creates challenging negative samples to enhance training. \cite{wang2024contrastive} combines contrastive learning with graph neural networks to learn both similarities in the latent space and dependency structures between different instance pairs in a self-supervised manner.

\cite{liu2024semi} proposes an improved deviation network with feature selection to address the need for extensive data in anomaly detection. However, relying on feature selection poses a risk of missing critical attributes, which could potentially lead to undetected intrusions. To address both the challenges of limited labeled data and privacy concerns in IIoT applications, \cite{aouedi2022federated} utilizes semi-supervised learning integrated with federated learning. Their architecture employs an autoencoder with a fully connected layer for intrusion detection, allowing the use of unlabeled data while preserving data privacy. The research in \cite{liu2024semi} introduces a semi-supervised approach incorporating information gain (IG) and principal component analysis (PCA) for feature selection, alongside a Gaussian-based deviation loss function. However, as noted before, feature selection in intrusion detection systems can be cumbersome.

Apart from learning-based methods, probabilistic and mathematical approaches have also been examined in the literature. One example is the work proposed in \cite{khalil2016novel}, which presents a probabilistically timed dynamic model for simulating physical security attacks on critical infrastructures. This model captures an attacker's behavior as they progress through vulnerabilities. It calculates mission success probability based on cumulative attack time relative to a given mission duration and employs Monte Carlo sampling to handle uncertainty. Additionally, it generates dynamic attack trees and visual flowcharts to better represent the decision-making process. 

In \cite{xiao2024multi}, colored Petri nets and mixed-strategy Nash equilibrium (CPNE) are used to predict multi-step attack paths for oil and gas intelligent pipelines. This method models asset importance and dependency relationships, simulates attack-defense interactions, and calculates node failure probabilities, from which multi-step attack paths are extracted.

\section{Datasets} \label{sec:Datasets}
In this section, we discuss the IoT datasets used in this research. Numerous publicly available IoT datasets exist in the literature, such as SWaT \cite{mathur2016swat}, WADI \cite{ahmed2017wadi}, Gas Pipeline \cite{morris2011control, morris2014industrial}, UNSW-NB15 \cite{moustafa2015unsw}, and BoT-IoT \cite{koroniotis2019towards}. 

There are two major types of datasets: simulated datasets, such as UNSW-NB15, and datasets collected from physical testbeds, such as SWaT, WADI, Gas Pipeline, and BoT-IoT. We validated our method using the UNSW-NB15, BoT-IoT, and Gas Pipeline datasets. These datasets primarily consist of network packet data. In the following, we provide a detailed discussion of each dataset.

\subsection{UNSW-NB15 Dataset}
The UNSW-NB15 dataset, created by researchers at the University of New South Wales (UNSW), represents modern network traffic scenarios and includes a wide range of low-footprint attacks. A synthetic testbed was configured using the IXIA PerfectStorm tool to generate realistic network traffic and diverse attack scenarios. This dataset contains network traffic collected from multiple sources, including benign communications, malicious attacks, and routine operations such as software updates. Raw network packets were captured using the \texttt{tcpdump} tool and stored in the PCAP file format.

A total of 49 network-based features were extracted across five categories: basic features, content features, time features, connection features, and additional features such as protocol type, duration, packet/byte counts, payload analysis, jitter, inter-packet arrival times, and historical behavior. This dataset includes multiple modern attack families, including Fuzzers, Analysis, Backdoors, DoS, Exploits, Reconnaissance, Shellcode, Generic, and Worms, in addition to Normal traffic. In this work, we utilize only a subset of categories: Normal, Fuzzers, DoS, Exploits, Generic, and Reconnaissance. The dataset contains approximately 2.5 million records, consisting of 2,218,761 normal and 321,283 attack records \cite{moustafa2015unsw}.

The UNSW-NB15 dataset was created to address the limitations of previous datasets (e.g., KDD99 and NSL-KDD), such as the lack of modern low-footprint attacks and outdated normal traffic patterns. Skewness and kurtosis tests on this dataset confirm non-linearity, non-normal distributions, and feature correlations that challenge traditional machine learning methods \cite{moustafa2016evaluation}. Older datasets like KDD99 and NSL-KDD suffered from redundant records, which UNSW-NB15 aims to overcome \cite{moustafa2017big,moustafa2017novel}. The various classes and the corresponding number of instances for each category are provided in \autoref{tab_unsw_nb15_dataset}.

It is worth noting that the original dataset contains 49 columns, including several categorical features: \textit{proto}, \textit{service}, and \textit{state}. We apply one-hot encoding to these features, resulting in a total of 204 columns after encoding. In this work, 558,954 samples are used for the pre-training phase and 55,895 samples for the fine-tuning stage, corresponding to 22.01\% and 2.20\% of the total number of samples, respectively. The fine-tuning set is sampled from the pre-training set.

\subsection{BoT-IoT Dataset}
The BoT-IoT dataset \cite{koroniotis2019towards}, publicly released in 2018 by the Network Security Lab at the University of New Brunswick, Canada, was collected using a smart home testbed that included various IoT devices such as routers, cameras, and smart appliances. The setup was designed to mimic real-world IoT environments and includes multiple types of cyber-attacks. Network traffic was recorded over several weeks using a passive monitoring tool called CICFlowMeter, which captured data at the network gateway.

This dataset contains network features such as protocol type, source and destination IP addresses, and port numbers, with each data point labeled. It serves as a benchmark for evaluating intrusion detection systems, particularly those focused on IoT security, and includes common IoT-related attacks. The categories present in this dataset are DDoS, DoS, Normal, Theft, and Reconnaissance.

In practice, when preparing the data for training, some columns such as \textit{attack} and \textit{subcategory} are removed, while the \textit{category} column is retained to serve as the target label. The \textit{proto} column, being categorical, is transformed using one-hot encoding, similar to the preprocessing steps applied to the UNSW-NB15 dataset. After encoding, the dataset contains 20 features for training. For this dataset, we use 469,570 samples, approximately 12.8\% of the total data, for the pre-training phase, while 46,957 samples, accounting for 1.28\% of the total 3,668,522 samples, are employed for the fine-tuning phase. The fine-tuning subset is sampled from the pre-training subset, and labels are used only for the fine-tuning subset. Both sample sizes are significantly lower than the volumes typically required by other algorithms. \autoref{tab_bot_iot_dataset} presents the classes in the BoT-IoT dataset along with the number of instances for each class.

\subsection{Gas Pipeline Dataset}
The Gas Pipeline dataset is developed to study intrusion detection in critical infrastructure systems that rely on SCADA technology. Its primary purpose is to simulate cyber-attacks targeting SCADA systems, serving as a standardized and common benchmark for evaluating different intrusion detection methods. This dataset is based on a small-scale gas pipeline testbed, where MODBUS traffic was recorded using a network data logger. The testbed consists of an airtight pipeline connected to a compressor, a pressure sensor, and a solenoid-controlled relief valve. The system regulates air pressure within the pipeline through a proportional-integral-derivative (PID) control scheme. In this research, we utilize the 2014 version of the dataset \cite{morris2014industrial}.

The dataset is composed of 27 columns, where the \textit{result} column denotes the condition of the system. Therefore, the dataset contains 26 actual features related to network traffic, process control, and process measurements. Among these, seven features contain only a single constant value and do not contribute useful information to the algorithm. These features are \textit{rate}, \textit{comm\_write\_fun}, \textit{gain}, \textit{reset}, \textit{deadband}, \textit{cycletime}, and \textit{crc\_rate}. After dropping these features, 19 features remain for training the network. Regarding the labels, there are eight distinct conditions in the dataset: normal (Normal), Naïve Malicious Response Injection (NMRI), Complex Malicious Response Injection (CMRI), Malicious State Command Injection (MSCI), Malicious Parameter Command Injection (MPCI), Malicious Function Code Injection (MDCI), Denial of Service (DoS), and Reconnaissance (Recon). These are denoted by labels 0 to 7, respectively \cite{morris2014industrial}. The classes along with their respective instance counts are shown in \autoref{tab_gas_pipeline_dataset}.

A total of 77,615 samples, representing 80\% of the entire dataset, are used for the pre-training phase. From this set, 7,761 samples (8\% of the total data) are selected for the fine-tuning phase. The pre-training stage does not require labels, which are only utilized during fine-tuning. This approach makes the method suitable for scenarios where abundant unlabeled data is available.

\begin{table}[] 
\caption{Number of data for UNSW-NB15 dataset}
\label{tab_unsw_nb15_dataset}
\centering
\setlength{\tabcolsep}{60pt} 
\begin{tabular}{@{}ll@{}}
\toprule
\textbf{Category} & \textbf{Number} \\ \midrule
Normal     & 677,785  \\
Fuzzers    & 5,051  \\
DoS        & 1,167  \\
Exploits   & 5,408 \\
Generic    & 7,522 \\
Recon      & 1,759 \\ \hline
Total      & 698,692\\ \bottomrule
\end{tabular}
\end{table}

\begin{table}[] 
\caption{Number of data for BoT-IoT dataset}
\label{tab_bot_iot_dataset}
\centering
\setlength{\tabcolsep}{60pt} 
\begin{tabular}{@{}ll@{}}
\toprule
\textbf{Category} & \textbf{Number} \\ \midrule
DDoS    & 1,926,624  \\
DoS     & 1,650,260  \\
Normal  & 477 \\
Theft   & 79 \\
Recon   & 91,082 \\ \hline
Total   & 3,668,522\\ \bottomrule
\end{tabular}
\end{table}

\begin{table}[] 
\caption{Number of data for Gas Pipeline dataset}
\label{tab_gas_pipeline_dataset}
\centering
\setlength{\tabcolsep}{60pt} 
\begin{tabular}{@{}ll@{}}
\toprule
\textbf{Category} & \textbf{Number} \\ \midrule
Normal & 61,156  \\ 
NMRI   & 2,763   \\
CMRI   & 15,466  \\ 
MSCI   & 782     \\
MPCI   & 7,637   \\ 
MDCI   & 573     \\
DoS    & 1,837   \\
Recon  & 6,805   \\ \hline
Total & 97,019 \\ \bottomrule
\end{tabular}
\end{table}

\section{Methodology}\label{sec:Method}
In this section, we present the methodology proposed in this study. Section \autoref{subsec_KAN} begins with an overview of the Kolmogorov-Arnold representation theorem and the KAN model. Then, \autoref{subsec_preprocessing} outlines the pre-processing steps associated with the method. Finally, \autoref{section_contrastive} provides a detailed discussion of the contrastive learning approach.

\subsection{KAN: Kolmogorov-Arnold Network}\label{subsec_KAN}
KAN, first introduced in \cite{liu2024kan}, is an alternative to traditional fully-connected networks. It exhibits strong capabilities in modeling non-linear functions and outperforms conventional MLP layers. KAN has demonstrated significant potential in a variety of domains, including anomaly detection \cite{abudurexiti2025explainable}, human activity recognition \cite{alikhani2025kan}, solving differential equations \cite{ma2024integrating}, and time series analysis \cite{lee2024hippo}. KAN has shown notable results such as advancements in accuracy, interpretability, and computational efficiency \cite{liu2024kan}. 

Before proceeding to the proposed methodology, we examine the current applications of KANs in the field of anomaly detection. \cite{ghorbani2025using} employs KAN for CPS attack detection using a minimal and fast approach that only utilizes standard scaling; this method achieves high performance with low inference time.

In \cite{mishra2025secure}, the authors propose faster-KAN with the help of Gaussian radial basis functions (GRBF) to detect anomalies in IoT networks. This research is suitable for low-end devices and systems with limited computational resources. \cite{abudurexiti2025explainable} introduces CCTAK, an unsupervised anomaly detection framework for IIoT networks, combining CNN and KAN. The enhanced interpretability of this method is a key advantage. \cite{yeonjeong2024robust} uses KAN for credit card fraud detection, where KAN outperforms conventional MLP networks. 

In \cite{do2025classifying}, the detection of botnet attacks is investigated. It is shown that the original KAN outperforms traditional ML and DL methods while also providing improved interpretability. \cite{del2024kolmogorov} utilizes KAN for motor condition monitoring, improving real-time diagnosis of faults in motors and achieving better performance and interpretability compared to MLP networks. KAN is also adopted in the healthcare domain. In \cite{huang2024abnormality}, the authors performed ECG-based cardiac abnormality detection, showcasing the effectiveness and generalizability of KAN for wearable medical devices.

KAN is grounded in the Kolmogorov-Arnold representation theorem, which initially served to solve differential equations. However, in this study, KAN is applied for classification tasks. The theorem guarantees that any multivariate continuous function $f(X)$, where $X = (X_1, \dots, X_{T})$, can be decomposed into a sum of nested univariate functions. Meaning, for the smooth differentiable function $f: [0, 1]^T \rightarrow \mathbb{R}^{d_{out}}$, there exist continuous univariate functions $\Phi_q$ and $\phi_{q,p}$, that satisfy the following equation:

\begin{equation}
f(X) = \sum^{2T+1}_{q=1}\Phi_q \bigl ( \sum^T_{p=1}\phi_{q,p}(X_p) \bigr )
\label{eq:KA_theorem}
\end{equation}
where $\phi_{q,p}: [0, 1] \rightarrow \mathbb{R}$ and $\Phi_q: \mathbb{R} \rightarrow \mathbb{R}$. In this formulation, each $\phi_{q,p}$ and $\Phi_q$ is a univariate function, and the nested sum of them enables the approximation of any continuous multivariate function. The Kolmogorov-Arnold representation theorem ensures that $(2T+1) \times T$ univariate functions are enough to approximate any given $T$-variate function. Compared to Taylor or Fourier series approximations, which require an infinite number of terms to represent a multivariate function with high precision, KANs offer a significant advantage by approximating a function using a finite composition of learnable univariate functions.

\begin{figure}
    \centering
    \includegraphics[width=0.5\linewidth]{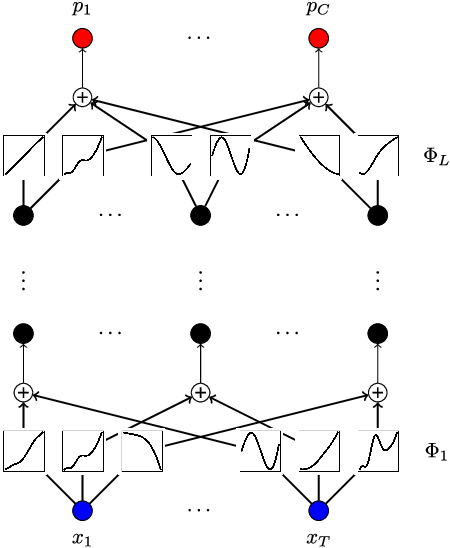}
    \caption{Visualization of a stack of $L$ KAN layers, with input dimension of $T$ and output dimension of $C$, denoting the number of classes \cite{baravsin2024exploring}.}
    \label{fig_KAN_architecture}
\end{figure}

KANs leverage this theoretical foundation by replacing typical pre-defined activation functions in neural networks with learned univariate functions applied to each input dimension individually. Unlike in a traditional MLP layer, where activation functions are applied to the outputs of nodes, KAN places activation functions on the edges of the network, i.e., between connected nodes. \autoref{fig_KAN_architecture} visualizes a KAN with an input size of $T$ and an output size of $C$. From this point on, to model the approximated function $f(\cdot)$, we use the notation $KAN(\cdot)$. Here, $X^i \in \mathbb{R}^T$ denotes a single sample, with a label $Y^i \in \{1, \dots, C\}$. In \autoref{fig_KAN_architecture}, the output shape is of dimension $C$. To formulate this more clearly, we can write $p = KAN(X^i)$ where $p = [p_1, p_2, \dots, p_C]$ is a vector of class representations.

The forward propagation of a KAN model with $L$ layers is expressed as a composition of non-linear transformations $\Phi_q$, each operating on the previous layer's output:
\begin{equation}
y = KAN(X) = (\Phi_L \circ \Phi_{L-1} \circ \dots \circ \Phi_{1})X
\label{eq:kan_in_matrix_form}
\end{equation}
each $\Phi_l$ consists of a set of learnable univariate functions $\phi$ applied element-wise. In contrast to traditional MLPs, where nonlinearity is introduced via fixed activation functions, KANs learn the activation functions during training. A traditional MLP network performs forward computation using linear weight matrices and fixed nonlinear activations:

\begin{equation}
y = MLP(X) = (W_L \circ \sigma \circ W_{L-1} \circ \sigma \circ \dots \circ W_{1})X
\label{eq_mlp_in_matrix_form}
\end{equation}
here, $W_l$ are weight matrices and $\sigma$ denotes a fixed activation function that introduces nonlinearity into the output of each layer. KANs learn the activation functions instead of using pre-defined ones. KANs use a hybrid activation function $\phi(X)$, which combines the SiLU and spline-based basis functions. The function is defined as follows:

\begin{equation}\label{eq:silu_spline}
\phi(X) = \omega_bSiLU(X)+\omega_sSpline(X)
\end{equation}
where the SiLU activation function is defined by:
\begin{equation}\label{eq:silu_activation}
SiLU(X) = \frac{X}{1 + e^{-X}}
\end{equation}

The SiLU activation function is a continuous function that guarantees the flow of gradients during the training process. This differentiable function has favorable optimization properties. The B-splines enable fine-grained modeling of nonlinearities through a learned basis function. Each spline function operates within a grid of size $G$:
\begin{equation}\label{eq:spline_basis_function}
Spline(X) = \sum ^{G+k}_{i=1} c_iB_i(X)
\end{equation}
here, $B_i(X)$ denotes a spline basis function, $c_i$ represents a learned coefficient or control point, $G$ is the number of grid points, and $K$ denotes the spline order (the default is $K=3$ for cubic splines). The spline order $K$ determines the smoothness of the curve, and a higher grid size $G$ increases the resolution of the splines. As shown in \autoref{eq:spline_basis_function}, a spline with order $K$ and grid size $G$ requires $G+K$ basis functions.

The total number of learnable parameters in a KAN layer is calculated as follows:

\begin{equation}\label{eq:KAN_parameters}
Parameters = (d_{in} \times d_{out})(G+K+3) + d_{out}
\end{equation}
here, $d_{in}$ and $d_{out}$ are the input and output dimensions, $G$ is the grid size or the number of grid points, and $K$ is the spline order. Our implementation uses two stacked KAN layers. The first layer acts as the feature extractor , and the second performs the final classification. Unlike traditional MLPs, where the parameter count depends only on the input and output dimensions, in KAN two additional parameters, the spline order and the grid size, affect the number of parameters. This results in more parameters even for equivalent dimensional configurations.

\subsection{Pre-processing}\label{subsec_preprocessing}The only pre-processing performed in this study is standardization using the mean ($\mu$) and standard deviation ($\sigma$) of the data. KAN has a unique ability to perform classification of intrusion data without requiring any additional pre-processing, resulting in faster inference times. The standardization is performed as follows:
\begin{equation}
X_{\text{scaled}} = \frac{X - \mu}{\sigma}
\label{eq_standard_scaler}
\end{equation}

We also experimented with min-max scaling; however, standardization outperformed min-max scaling in our experiments.

\begin{figure*}
    \centering
    \includegraphics[width=0.6\linewidth]{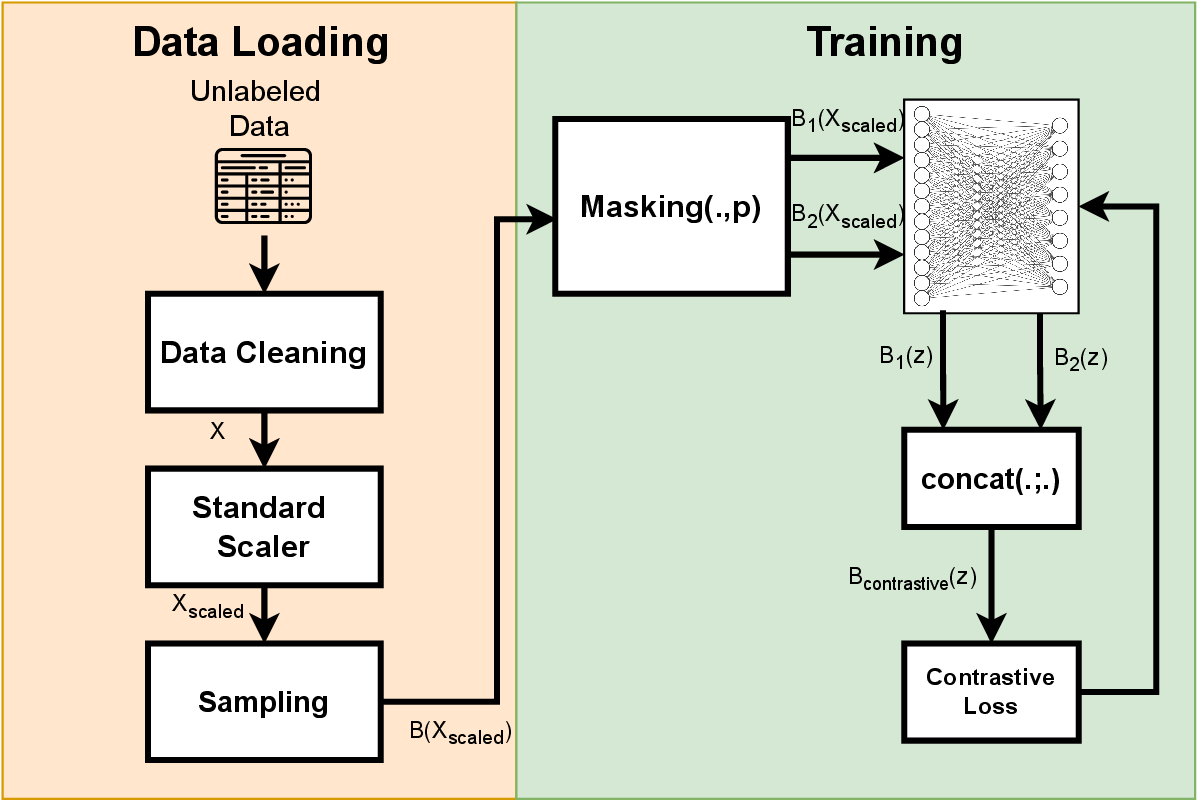}
    \caption{The contrastive pre-training framework. The data loading part involves, removing the unnecessary columns, encoding categorical features, handling NaN values, and sampling a batch of data for contrastive learning.}
    \label{fig_pre_training_framework}
\end{figure*}

\subsection{Contrastive learning}\label{section_contrastive}
The contrastive learning framework was first introduced in \cite{chen2020simple}, where it was applied in a self-supervised manner for visual representation learning. One of the key advantages of this method is its ability to utilize large amounts of unlabeled data. Such unlabeled data is common across various domains, which is a significant benefit since annotating data requires substantial time, domain expertise, and financial resources. Contrastive learning offers a scalable solution for leveraging large volumes of unlabeled data by pre-training a feature extractor without the need for labels, thereby learning the underlying features present in the data. After pre-training, the parameters of the feature extractor are frozen, and a final classifier is appended to the network. Fine-tuning is then performed only on this last layer using a small fraction of labeled data, without modifying the parameters of the FE.

The process begins with the data $D=\{X_\text{scaled}, Y\}$ where $X_\text{scaled} \in \mathbb{R}^{d_T}$ is standardized data, and also labels $Y$ are assumed to be unknown. Contrastive learning aims to learn similar representations for different views of the same data point that has gone through different data augmentations. 

Since similarity learning in contrastive training is based on intra-batch distances, the first step involves sampling batches of size $M$ from the data $X_{\text{scaled}}$. Following this procedure results in $B(X_{\text{scaled}})$, where the notation $B(\cdot)$ denotes a batch.

Now, we define the feature extractor, denoted by $KAN_{\theta}(\cdot)$ with learnable parameters $\theta$, which takes an input data point $X$ and generates its representation $z$. As mentioned earlier, contrastive learning requires different views of the same data point. To achieve this, we propose a masking-based data augmentation strategy that randomly sets some features in the data point $X$ to zero. This masking function is defined as:
\begin{equation}
    X_{\text{masked}} = \text{Masking}(\cdot,p)
    \label{masking_func}
\end{equation}
where $p$ is a pre-defined masking percentage, which is set to $0.1$ in our case. The masking function randomly sets $p\%$ of the input features to zero. By applying the masking function described in \autoref{masking_func} to an input batch twice, we obtain two different augmented versions of the same input batch $B(X_{\text{scaled}})$, denoted by $B_1(X)$ and $B_2(X)$. For simplicity, we omit the \textit{scaled} subscript, which previously indicated that the data was standardized. 

As mentioned earlier, the feature extractor $KAN_\theta(\cdot)$ should produce similar representations for corresponding elements in these two augmented batches. Let $B_1(z)$ and $B_2(z)$ represent the resulting batches of learned representations obtained from each view, where $z$ denotes the output of the feature extractor. This process can be expressed as follows:

\begin{align}
    B_1(z) = KAN_{\theta}(B_1(X))\\
    B_2(z) = KAN_{\theta}(B_2(X))
\end{align}
where the network $KAN_{\theta}(\cdot)$ generates a batch of representations for each input batch. Both the input and output batches have a size of $M$. The objective is to bring the representations corresponding to the same original data point (across different augmentations) closer together in the representation space, while pushing apart the representations of dissimilar (non-matching) data points.

To achieve this, the two augmented batches are concatenated to form a single batch of size $2M$, as follows:
\begin{equation}
    B_{\text{contrastive}}(z) = concat(B_1(z);B_2(z))
\end{equation}

The contrastive loss utilized in this research is based on cosine similarity:
\begin{equation}
    l_{i,j} = \log \frac{\exp(\text{sim}_{\text{cos\_sim}}(x^i, x^j)/\tau)}{\sum_{k=1}^{2M} 1_{k \ne i} \exp(\text{sim}_{\text{cos\_sim}}(x^i, x^k)/\tau)}
    \label{eq_contrastive_loss_fn}
\end{equation}
where $i,j \in\{1, \dots, 2M\}$, denote the indices of the samples in batch $B_{\text{contrastive}}(z))$ and $\text{sim}_{\text{cos\_sim}}$ is the cosine similarity between two vectors as follows:
\begin{equation}
    \text{sim}_{\text{cos\_sim}} = \cos(\phi) = \frac{\mathbf{A} \cdot \mathbf{B}}{\|\mathbf{A}\| \|\mathbf{B}\|}
\end{equation}
where $\phi$ is the angle between the two latent representations. In \autoref{eq_contrastive_loss_fn}, the self-similarity ($i = k$) is ignored using the indicator function $1_{k \ne i}$. The term $\tau$ is the temperature (or scaling) factor, which is set to $0.5$. The notation $x^i = B_{\text{contrastive}}(z)(i)$ represents the $i^{\text{th}}$ element of the batch $B_{\text{contrastive}}(z)$, with similar definitions for $x^j$ and $x^k$.

$l$ is a $2M \times 2M$ matrix, where the elements $l(2k, 2k - 1)$ and $l(2k - 1, 2k)$ represent the pairwise similarities between the two augmented batches. The total loss is then calculated as:
\begin{equation}
    Loss = \frac{1}{2M} \sum_{k=1}^{M} (l_{2k-1, 2k} + l_{2k, 2k-1})
    \label{contrastive_loss_for_class_c}
\end{equation}

Since the denominator of \autoref{eq_contrastive_loss_fn} includes all the dissimilar samples, it captures the dissimilarity associated with the negative samples. Maximizing the distance between dissimilar samples results in minimizing the loss. The parameters $\theta$ of the feature extractor are updated using gradient descent optimization, as follows:

\begin{equation}
    \theta_{n} = \theta_{n-1} - \eta \frac{\partial Loss}{\partial \theta}
\end{equation}
where $\eta$ is the pre-training learning rate, and $n$ denotes the $n^{th}$ iteration of training. The weights are then saved for the fine-tuning stage. Algorithm \autoref{alg_contrastive_kan} and Figure \autoref{fig_pre_training_framework} present an algorithmic and visual overview of the contrastive training process, respectively.

\begin{algorithm} 
\caption{Proposed contrastive learning framework}
\begin{algorithmic}[1]
\REQUIRE 
    \textit{epochs}: number of epochs, \\
    \textit{Input data}: $X$ \\
    \textit{Masking}: Making a function that masks a percentage of features randomly, \\
    \textit{$p$}: Masking percentage \\
    \textit{$\eta$}: Learning rate\\
    \textit{$KAN_{\theta}(\cdot)$}: The feature extractor with learnable parameters $\theta$\\
    \STATE Normalize data: $X_{\text{scaled}} \gets \frac{X - \mu}{\sigma}$
\FOR{$epoch = 1$ to $epochs$}
    \STATE Create a batch of training data, with size $M$: $B(X_{\text{scaled}})$ 
    \STATE Create different views of the data with augmentations:\\
    $B_1(X) \gets \text{Masking}(B(X_{\text{scaled}}), p)$\\
    $B_2(X) \gets \text{Masking}(B(X_{\text{scaled}}), p)$\\
    \STATE Feed each of the batches to the model $KAN_{\theta}(\cdot)$:
    $B_1(z) \gets KAN_{\theta}(B_1(X))$\\
    $B_2(z) \gets KAN_{\theta}(B_2(X))$\\
    \STATE Concatenate the two batches of representations vertically to create a batch of $2M$ samples:\\
    $B_{\text{contrastive}}(z) \gets concat(B_1(z);B_2(z))$\\
    \STATE Calculate the $l_{i,j}$ with the contrastive loss, and $B_{\text{contrastive}}(z)$
    \STATE Compute similarity loss by adding the elements of the $(2k-1, 2k)$ and $(2k, 2k-1)$:\\
    $Loss \gets \frac{1}{2M} \sum_{k=1}^{M} (l_{2k-1, 2k} + l_{2k, 2k-1})$
    \STATE Update feature extractor's parameters: \\
    $\theta_{n} \gets \theta_{n-1} - \eta \frac{\partial Loss}{\partial \theta}$\\
    \STATE Save model $KAN_{\theta}(\cdot)$ 
    \STATE Use $KAN_{\theta}(\cdot)$ to evaluate new data and use for fine-tuning
\ENDFOR
\end{algorithmic}
\label{alg_contrastive_kan}
\end{algorithm}

\subsection{Fine-tuning}
After pre-training the $KAN_\theta(\cdot)$, we stack another KAN layer, denoted as $KAN_\lambda(\cdot)$, with learnable parameters $\lambda$, to form the final classifier. From this point onward, batch notation is omitted for simplicity. An important aspect of the training process is that only the new layer parameters, denoted by $\lambda$, are updated, while the pre-trained parameters $\theta$ remain frozen and are not updated during the fine-tuning stage. The labeled dataset used during fine-tuning is denoted as $D = (X_{\text{scaled}}, Y)$, where $X_{\text{scaled}}$ is the standardized input feature vector and $Y^i \in \{0, \dots, C\}$ is the label. Unlike in the pre-training phase, where labels were unknown, the fine-tuning stage uses the labels during training. The notation for the $i^{th}$ data point is defined as:
\begin{equation}
    z^i=KAN_\theta(X_{\text{scaled}}^i)\label{eq_latent_space_rep}
\end{equation}
\begin{equation}
     \zeta^{(i)} = KAN_{\lambda}(z^i)\label{eq_class_rep}
\end{equation}
\begin{equation}
    class^{(i)} = \mathrm{softmax}(\zeta^{(i)}) \label{eq_class_softmax}
\end{equation}
\begin{equation}
    \text{error}_\text{final} = \sum_{i=1}^{N} \text{crossEntropy}(class^{(i)}, Y^{(i)}) \label{eq_cross_entropy}
\end{equation}

First, we obtain the latent space representations from the feature extractor $KAN_\theta(\cdot)$, as defined in \autoref{eq_latent_space_rep}. These representations are then passed through the newly stacked KAN layer $KAN_\lambda(\cdot)$ to produce the class representations, based on \autoref{eq_class_rep}. The resulting class representations are fed into the softmax function to generate class probabilities. Next, these probabilities are used in the cross-entropy loss function, shown in \autoref{eq_cross_entropy}, which computes the final classification loss, $\text{error}_\text{final}$. This loss is then employed in a gradient descent algorithm to update the model parameters as follows:
\begin{equation}
    \lambda_{n+1} = \lambda_n - \alpha \frac{\partial \text{error}_\text{final}}{\partial \lambda}
\end{equation}
where $\alpha$ is the fine-tuning learning rate, and $\lambda$ denotes the parameters of the final layer. After completing the training process, the model can be saved and used for inference on new data. Algorithm \autoref{alg_fine_tuning} presents the pseudo-code for the fine-tuning phase of the proposed method.

\begin{algorithm}
\caption{Fine-tuning of the proposed intrusion detection framework}
\begin{algorithmic}[1]
\REQUIRE
    \textit{epochs}: number of epochs, \\
    \textit{Input data}: $(X, Y)$ \\
    \textit{$\alpha$}: Learning rate\\
    \textit{$KAN_{\theta}(\cdot)$}: The feature extractor with parameters $\theta$\\
    \textit{$KAN_{\lambda}(\cdot)$}: The feature extractor with learnable parameters $\lambda$\\
    \STATE Normalize with the $\mu$ and $\sigma$ derived from pre-training:\\ 
    $X_{\text{scaled}} \gets \frac{X - \mu}{\sigma}$\\
    \FOR{each parameters $\in KAN_{\theta}$}
        \STATE parameters$.requires\_grad \gets False$
    \ENDFOR
\FOR{$epoch = 1$ to $epochs$}
    \STATE Feed the data point $X^i_{scaled}$ to $KAN_{\theta}(\cdot)$:\\
    $z^i \gets KAN_{\theta}(X_{\text{scaled}}^i)$\\
    \STATE Get class representations:\\
    $\zeta^{(i)} \gets KAN_{\lambda}(z^i)$
    \STATE Class probabilities: $class^{(i)} \gets \mathrm{softmax}(\zeta^{(i)})$
    \STATE Compute  classification error: \\
    $\text{error}_\text{final} \gets \sum_{i=1}^{N} \text{crossEntropy}(class^{(i)}, Y^{(i)})$\\
    \STATE Update final KAN layer $KAN_{\lambda}(\cdot)$ parameters:\\
    $\lambda_{n} \gets \lambda_{n-1} - \alpha \frac{\partial \text{error}_\text{final}}{\partial \lambda}$\\
    \STATE Save model $KAN_{\lambda}$ 
    \STATE Use $KAN_{\lambda}$ in conjunction with $KAN_{\theta}$, to evaluate new data and use for fine-tuning
\ENDFOR
\end{algorithmic}
\label{alg_fine_tuning}
\end{algorithm}

\subsection{Simulation Setup} \label{sec:setup}
The implementation of this project is based on the \textit{efficient-kan} implementation\footnote{\url{https://www.github.com/Blealtan/efficient-kan}}. In these experiments, we utilize a learning rate of 0.001 for both the pre-training and fine-tuning stages across all datasets. The contrastive pre-training batch sizes are 64, 16, and 64 for the UNSW-NB15, BoT-IoT, and Gas Pipeline datasets, respectively. Likewise, the fine-tuning batch sizes are 64, 32, and 2, respectively. The masking percentage is set to $10\%$ for the UNSW-NB15 dataset and $20\%$ for the BoT-IoT and Gas Pipeline datasets. The higher masking percentage for the BoT-IoT and Gas Pipeline datasets is due to their lower feature dimensionality; with a smaller $p$, there is a higher chance of generating similar augmented views after masking. The Adam optimizer is used during the pre-training phase, while AdamW is employed for the fine-tuning stage. The experiments are performed in the Kaggle environment, which has 32 GB of RAM, an NVIDIA Tesla P100 GPU with 16 GB of VRAM, and an Intel(R) Xeon(R) CPU at 2.00 GHz.

\section{Results} \label{sec_Results}
\begin{figure*}[h!]
    \centering
    \begin{tabular}{ccc}
        \includegraphics[width=0.3\textwidth]{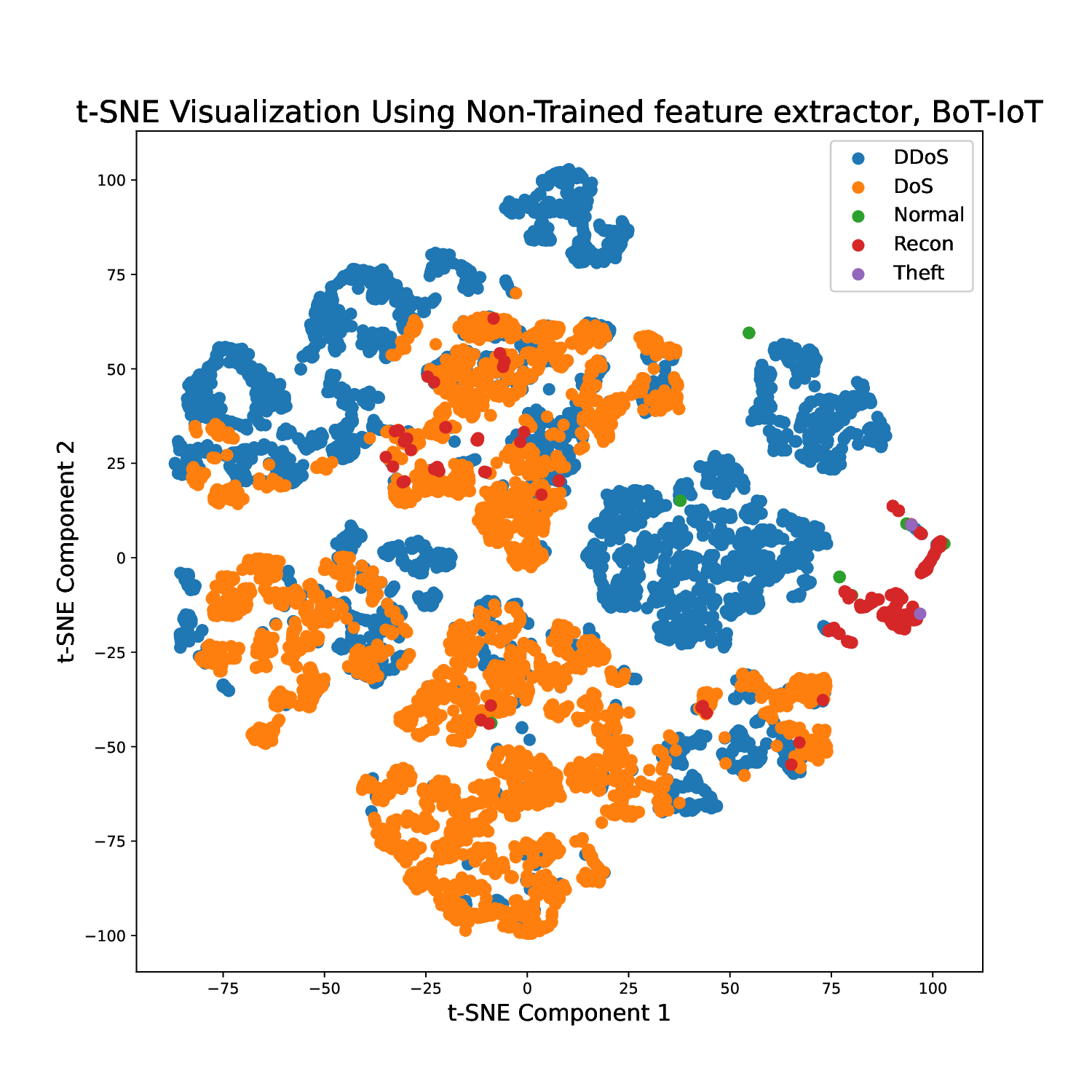} &
        \includegraphics[width=0.3\textwidth]{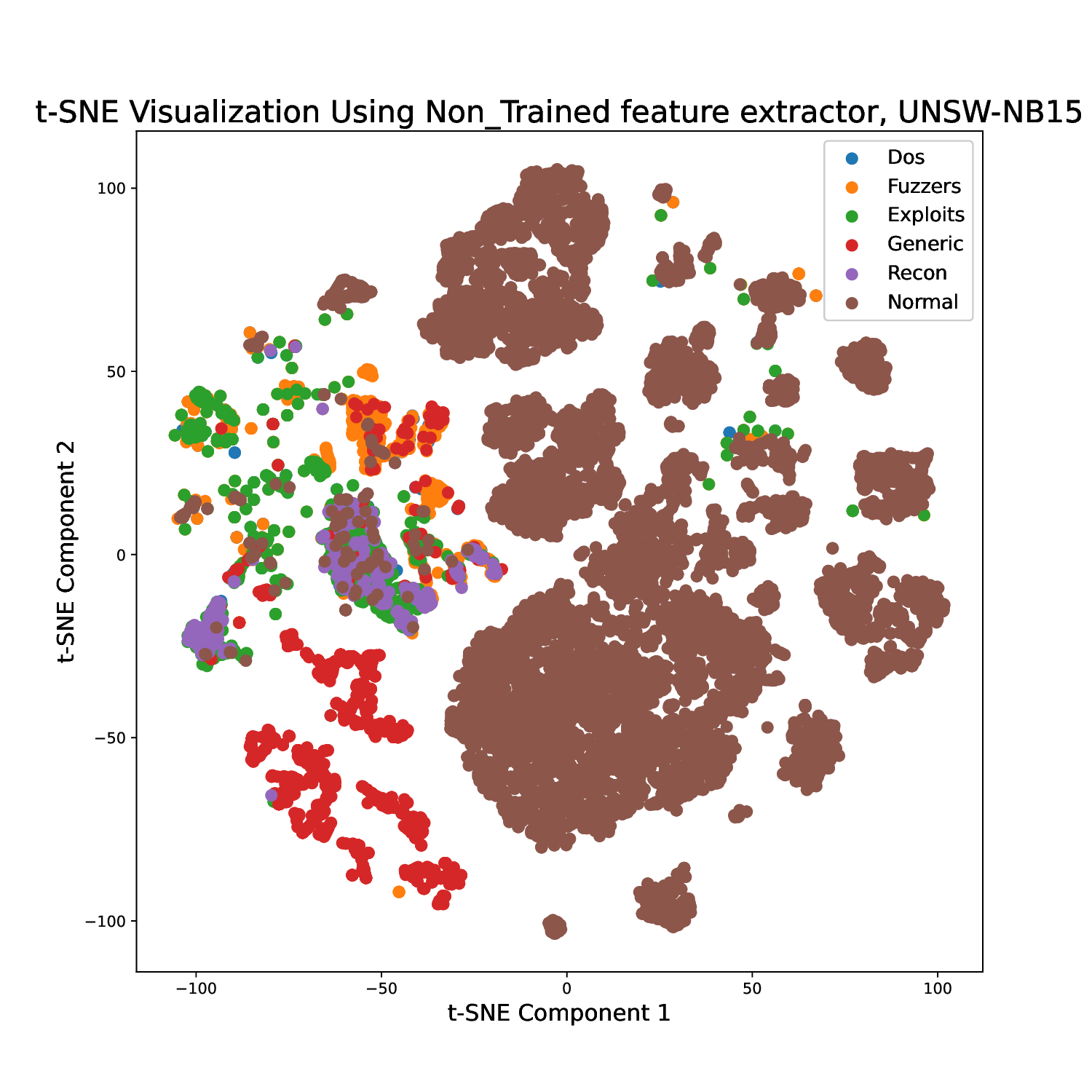} &
        \includegraphics[width=0.3\textwidth]{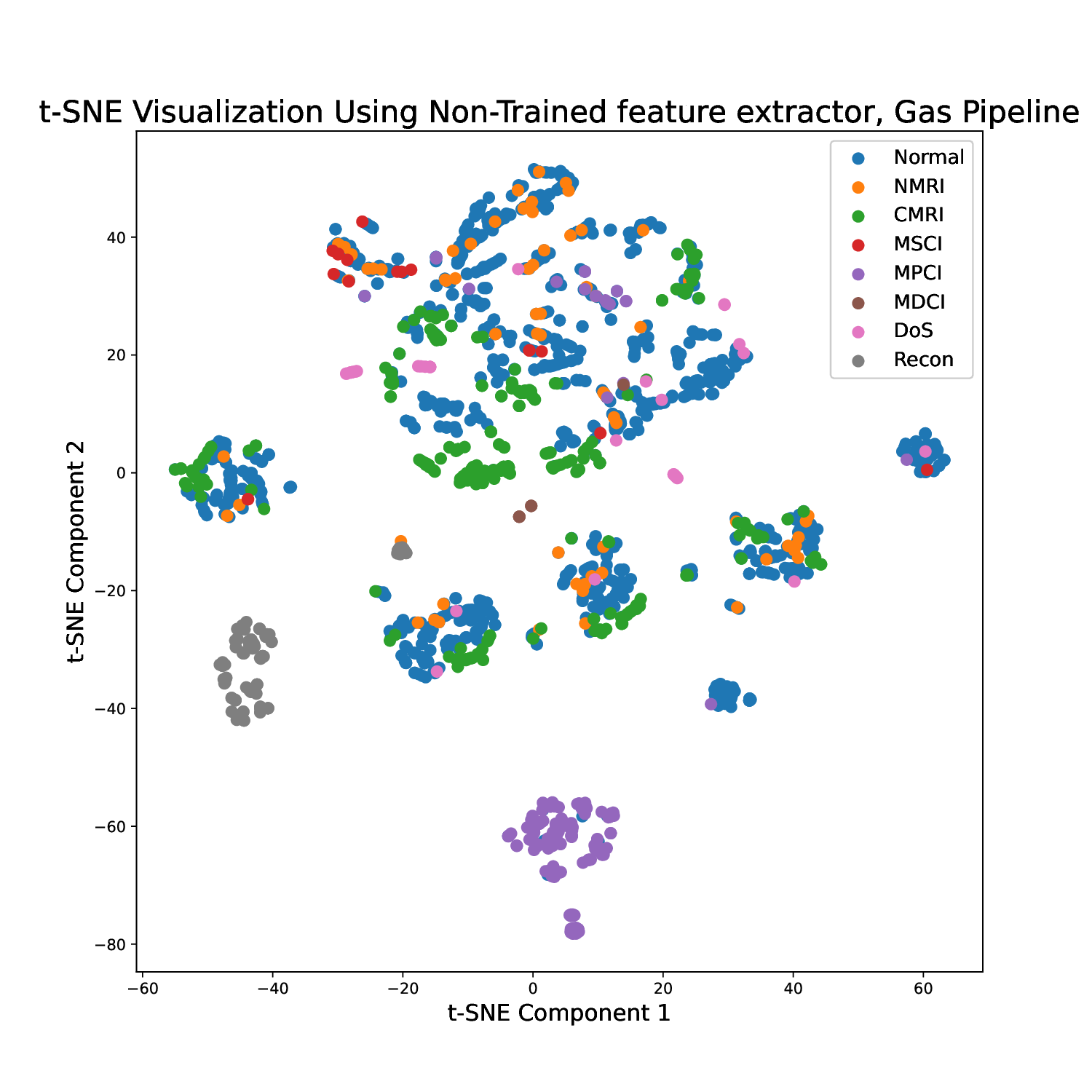}\\
        (a) BoT-IoT  & (b) UNSW-NB15 & (c) Gas Pipeline \\
    \end{tabular}
    \caption{t-SNE visualization of the feature extractor outputs, before pre-training.}
    \label{fig_tsne_before_train}
\end{figure*}

\begin{figure*}[h!]
    \centering
    \begin{tabular}{ccc}
        \includegraphics[width=0.3\textwidth]{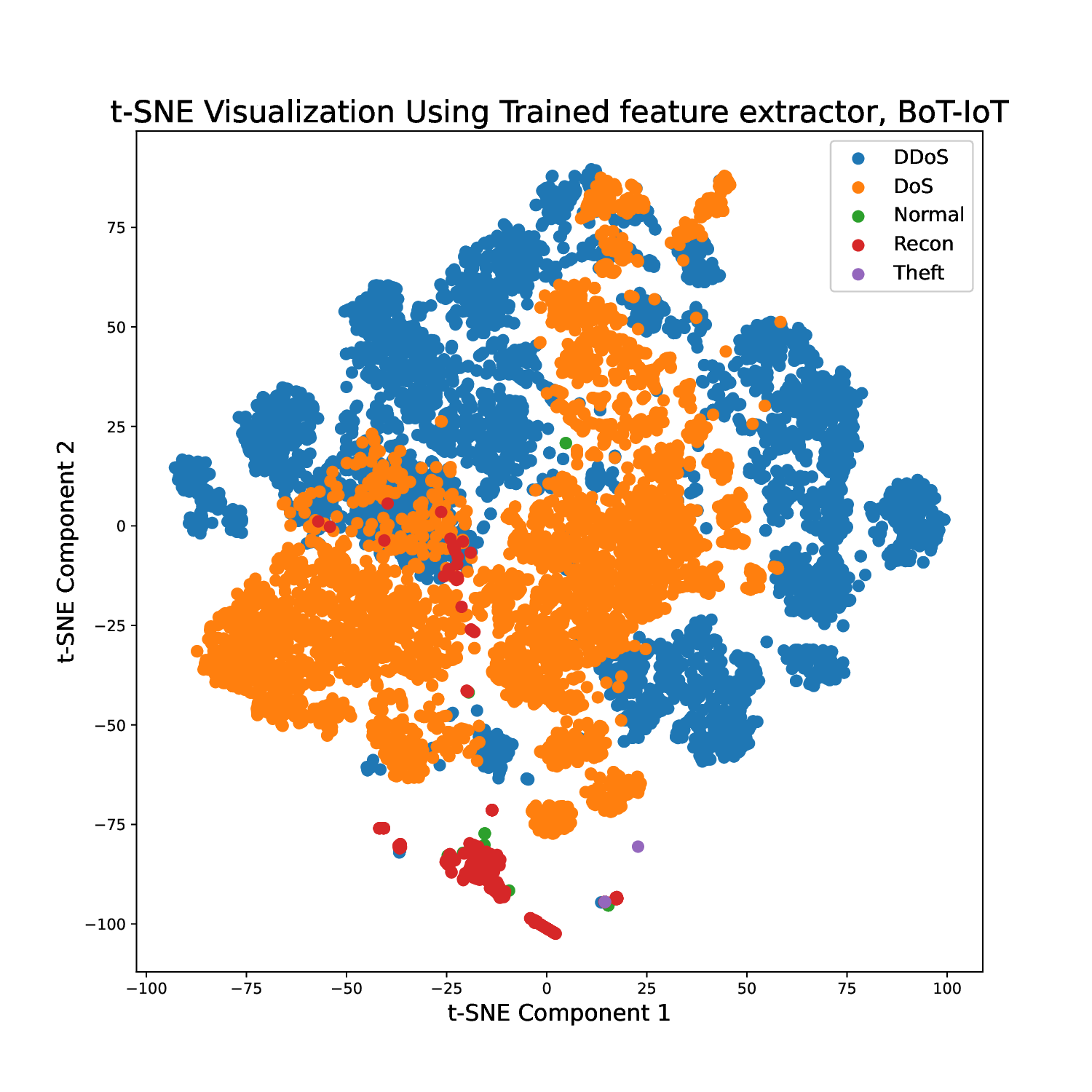} &
        \includegraphics[width=0.3\textwidth]{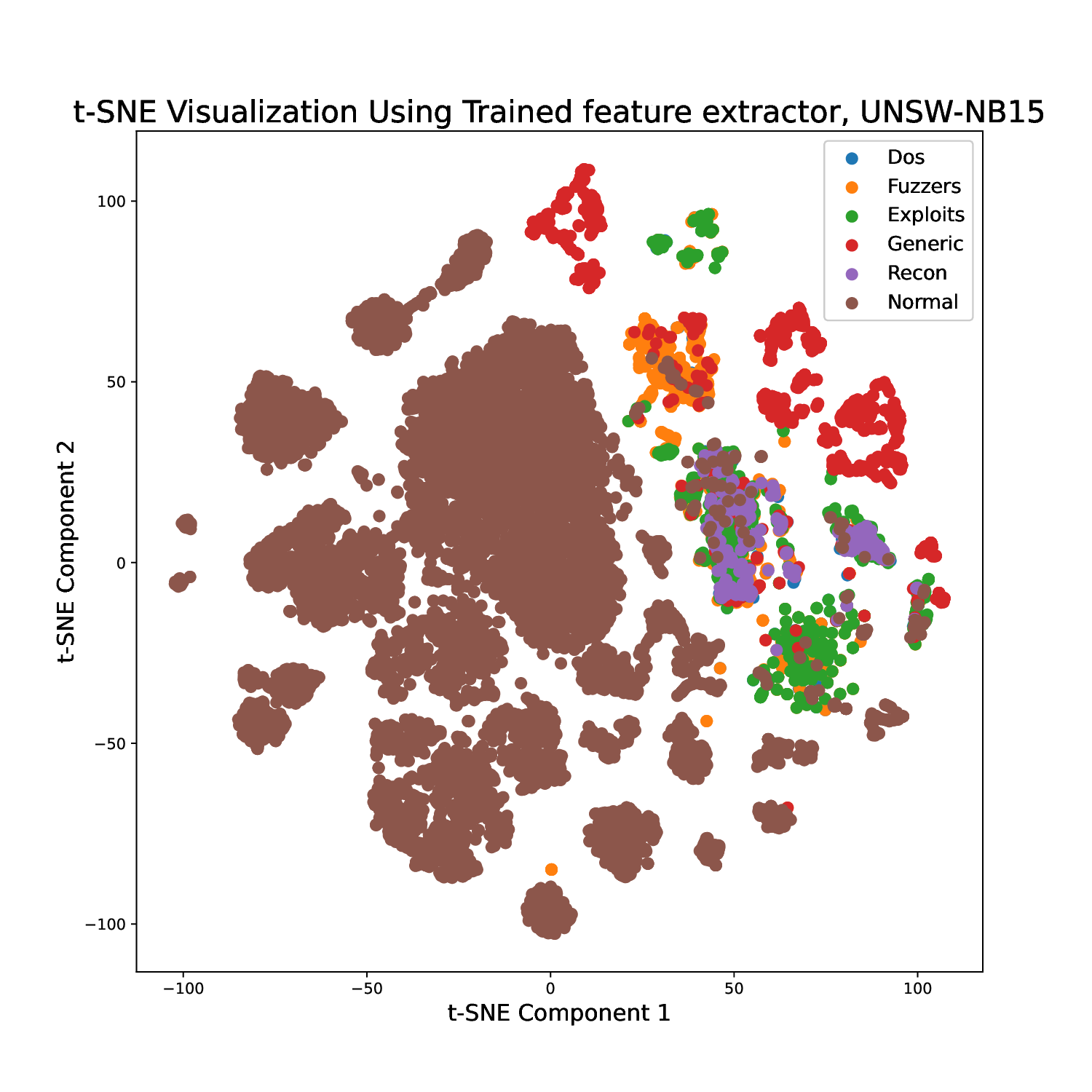} &
        \includegraphics[width=0.3\textwidth]{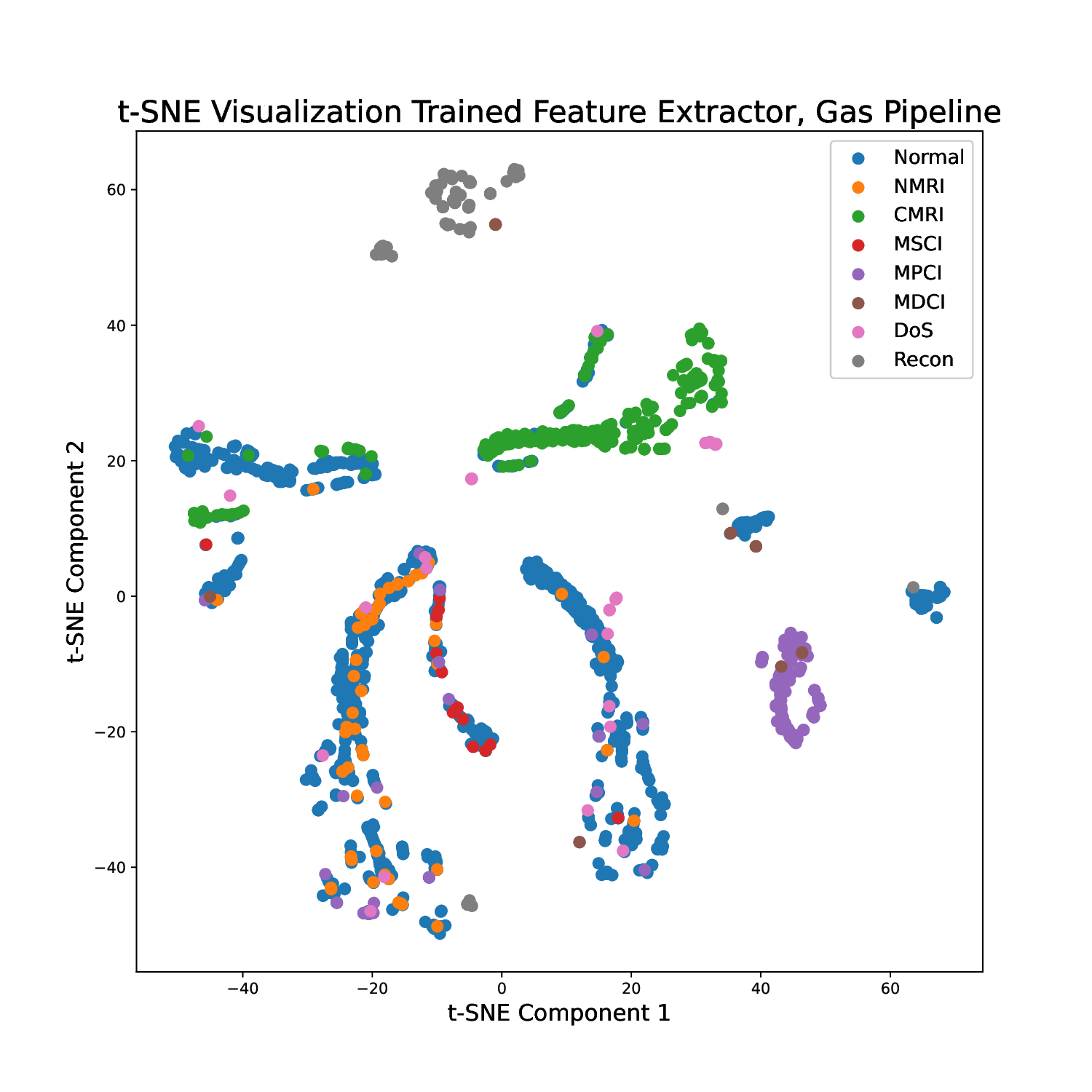}\\
        (a) BoT-IoT  & (b) UNSW-NB15 & (c) Gas Pipeline \\
    \end{tabular}
    \caption{t-SNE visualization of the feature extractor outputs, after pre-training.}
    \label{fig_tsne_after_train}
\end{figure*}

\subsection{Performance Analysis}
In this section, we analyze the results of the experiments conducted on the three different datasets described earlier. \autoref{fig_tsne_before_train} presents the t-SNE visualizations of the test data passed through the KAN feature extractor before pre-training, while \autoref{fig_tsne_after_train} shows the corresponding results after completing the pre-training. These figures demonstrate the effectiveness of the proposed contrastive learning framework.

In the BoT-IoT dataset, the DoS and DDoS samples form more compact clusters and exhibit reduced overlap after the network has been pre-trained. In the UNSW-NB15 dataset, following pre-training, the Exploits and Normal samples demonstrate a lower degree of overlap, resulting in better separability between Normal and attack samples. Additionally, the Exploits and Fuzzers data points are more distinctly clustered compared to the non-pretrained feature extractor. The t-SNE visualizations for the Gas Pipeline dataset clearly confirm the positive effect of contrastive pre-training on the separability of the test data. However, the Gas Pipeline dataset remains the most challenging among the three, due to its complex structure, smaller data volume, and fewer distinguishable features.
\begin{figure*}
    \centering
    \includegraphics[width=1\linewidth]{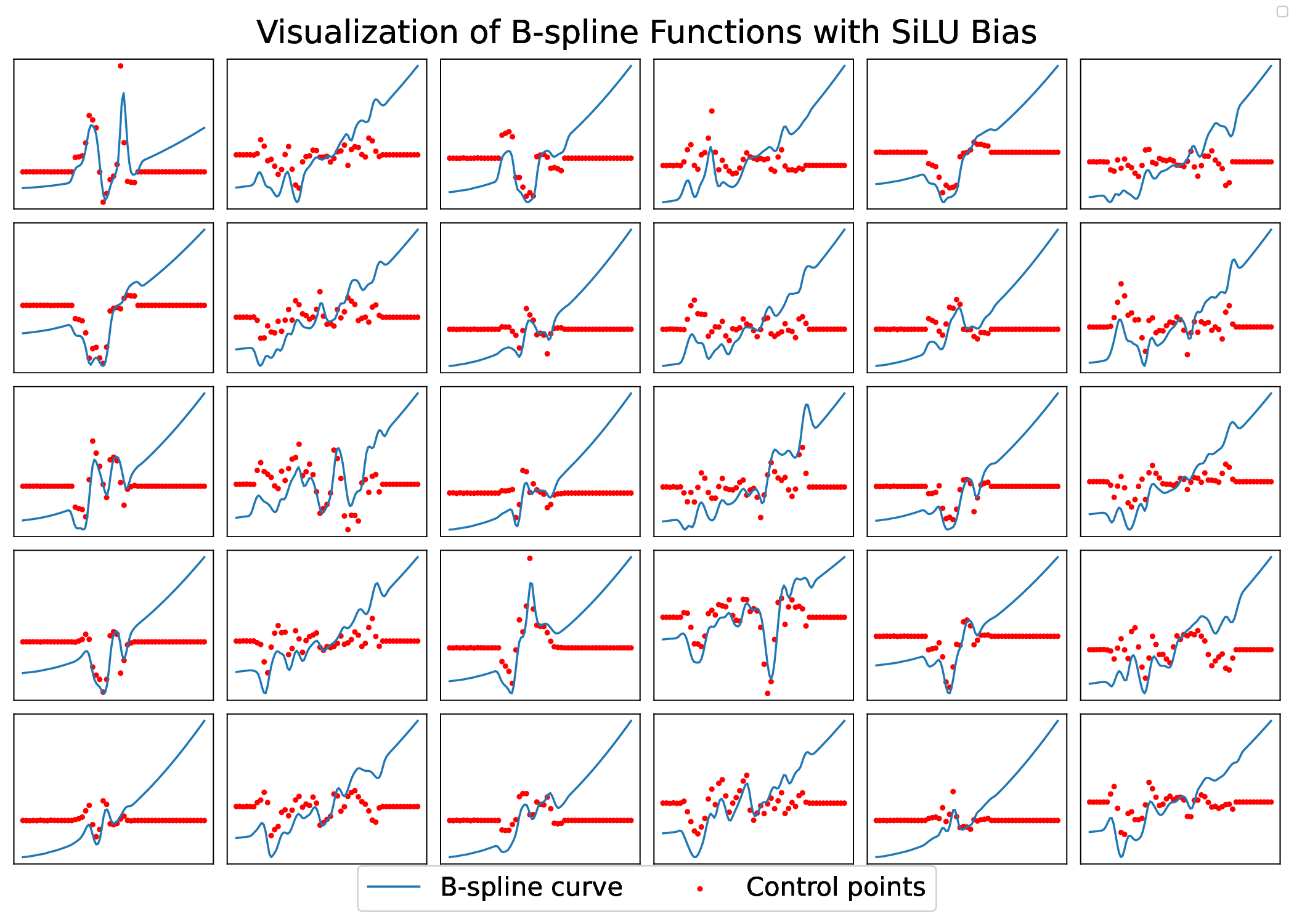}
    \caption{Visualization of the spline-based learned activation functions, aggregated with SiLU, for the UNSW-NB15 dataset. The red points represent the control points referenced in \autoref{eq:spline_basis_function}. The observed upward trend in the curves results from the addition of the SiLU function to the spline components, as described in \autoref{eq:silu_spline}.}
    \label{fig_spline_visualization}
\end{figure*}

\subsection{Inference Time}
We measured the inference time of the network to assess its feasibility for real-time deployment across all datasets. The proposed contrastive-KAN model was evaluated on the entire test set, and the average inference time per sample was reported. Specifically, the inference times were 0.41 ms for the Gas Pipeline dataset, 0.26 ms for the UNSW-NB15 dataset, and 0.13 ms for the BoT-IoT dataset. All evaluations were conducted on a CPU in the Kaggle environment, using the batch sizes specified during the fine-tuning stage. It is worth noting that smaller batch sizes introduce some computational overhead, as they require more iterations to process the entire dataset. Despite this, the measured inference times clearly demonstrate that our proposed method operates in real-time, providing fast and timely intrusion detection.

\subsection{Model Interpretability}
Given the critical need for safe and reliable operation in IIoT and IoT networks, explainable artificial intelligence (XAI) and model interpretability have become essential. In this context, KAN offers a valuable advantage by enabling the visualization of the learned activation functions. This feature allows engineers to extract meaningful rules from the trained intrusion detection system, which can subsequently be incorporated into rule-based approaches. Such interpretability further enhances the reliability and safety of these systems, particularly in safety-critical applications.

In \autoref{fig_spline_visualization}, the spline functions learned by the pre-trained feature extractor on the UNSW-NB15 dataset, biased by the SiLU activation function, are visualized. The red dots represent the control points, as discussed in \autoref{eq:spline_basis_function}. The observed upward trend in the curves results from the addition of the SiLU activation function. Although many activation functions are learned, visualizing all of them is impractical. Nonetheless, \autoref{fig_spline_visualization} effectively demonstrates the feasibility of this interpretability feature.

\begin{table}[ht]
\caption{F1-scores on the UNSW-NB15 dataset for various combinations of spline orders and grid sizes, obtained by training with the proposed framework. The network architecture consists of two layers with dimensions [204, 50, 6].}
\label{tab_UNSW_15B_f1_grid_contrastive}
\centering
\begin{tabular}{@{}l|l|cccc@{}}  
\toprule
\multicolumn{2}{c}{} & \multicolumn{4}{c}{\textbf{Grid Size}} \\ \cmidrule(l){3-6}
\multicolumn{2}{c}{} & \textbf{1} & \textbf{5} & \textbf{20} & \textbf{50} \\ \midrule
\multirow{4}{*}{\textbf{Spline Order}} 
& \textbf{1}   & 82.03 & 84.08 & 89.43 & \textbf{93.66} \\
& \textbf{3}   & 83.14 & 87.45 & 90.17 & 91.24          \\
& \textbf{10}  & 84.01 & 86.25 & 87.35 & 89.31          \\
& \textbf{30}  & 83.12 & 85.14 & 87.29 & 88.27          \\ \bottomrule
\end{tabular}
\end{table}

\begin{table}[ht]
\caption{F1-scores on the BoT-IoT dataset for various combinations of spline orders and grid sizes, obtained by training with the proposed framework. The network architecture consists of two layers with dimensions [20, 100, 2].}
\label{tab_BoT_IoT_f1_grid_contrastive}
\centering
\begin{tabular}{@{}l|l|cccc@{}}  
\toprule
\multicolumn{2}{c}{} & \multicolumn{4}{c}{\textbf{Grid Size}} \\ \cmidrule(l){3-6}
\multicolumn{2}{c}{} & \textbf{1} & \textbf{5} & \textbf{20} & \textbf{50} \\ \midrule
\multirow{4}{*}{\textbf{Spline Order}} 
& \textbf{1}   & 85.12 & 89.02 & 91.43 & 92.66          \\
& \textbf{3}   & 86.21 & 88.99 & 91.00 & \textbf{93.11} \\
& \textbf{10}  & 86.74 & 88.53 & 92.19 & 92.22          \\
& \textbf{30}  & 87.44 & 88.51 & 90.20 & 91.16          \\ \bottomrule
\end{tabular}
\end{table}

\begin{table}[ht]
\caption{F1-scores on the Gas Pipeline dataset for various combinations of spline orders and grid sizes, trained using the proposed framework. The network consists of two layers with input dimensions [19, 100, 8].}
\label{tab_gas_pipeline_f1_grid_contrastive}
\centering
\begin{tabular}{@{}l|l|cccc@{}}  
\toprule
\multicolumn{2}{c}{} & \multicolumn{4}{c}{\textbf{Grid Size}} \\ \cmidrule(l){3-6}
\multicolumn{2}{c}{} & \textbf{1} & \textbf{5} & \textbf{20} & \textbf{50} \\ \midrule
\multirow{4}{*}{\textbf{Spline Order}} 
& \textbf{1}   & 86.15 & \textbf{86.70} & 86.32 & 86.10 \\
& \textbf{3}   & 86.18 & 86.66          & 86.33 & 86.02 \\
& \textbf{10}  & 86.27 & 85.86          & 86.33 & 86.17 \\
& \textbf{30}  & 86.18 & 85.97          & 86.61 & 86.53 \\ \bottomrule
\end{tabular}
\end{table}

\begin{table}[] 
\caption{Ablation study on the UNSW-NB15 dataset using the specified data split. The grid size and spline order are selected based on the best results reported in \autoref{tab_UNSW_15B_f1_grid_contrastive}.}
\label{tab_unsw_ablation}
\centering
\begin{tabular}{@{}llll@{}}
\toprule
\textbf{Model} & \textbf{Architecture}     & \textbf{\begin{tabular}{l} Num. of \\ Params \end{tabular}} & \textbf{F1 (\%)} \\ \midrule
KAN (end-to-end)   & {[}204 50 6{]}        & 605,472                  & 86.78                 \\
KAN (end-to-end)   & {[}204 70 6{]}        & 779,100                  & 87.44                  \\
KAN (end-to-end)   & {[}204 100 6{]}       & 1,113,000                & 85.65                  \\ \midrule
\textbf{KAN (contrastive)}  & \textbf{{[}204 50 6{]}}       & \textbf{605,472}                & \textbf{93.66}         \\
KAN (contrastive)  & {[}204 70 6{]}        & 779,100                  & 92.22                  \\
KAN (contrastive)  & {[}204 100 6{]}       & 1,113,000                & 91.91                  \\ \midrule
MLP (end-to-end)   & {[}204 600 800 6{]}   & 608,606                  & 83.59                  \\
MLP (end-to-end)   & {[}204 700 900 6{]}   & 779,806                  & 84.92                 \\
MLP (end-to-end)   & {[}204 940 990 6{]}   & 1,130,236                & 85.15                  \\ \midrule
MLP (contrastive)  & {[}204 600 800 6{]}   & 608,606                  & 87.25                  \\ 
MLP (contrastive)  & {[}204 700 900 6{]}   & 779,806                  & 88.87                  \\ 
MLP (contrastive)  & {[}204 940 990 6{]}   & 1,130,236                & 88.98                  \\ \bottomrule
\end{tabular}
\end{table}

\begin{table}[] 
\caption{Ablation study on the BoT-IoT dataset using the specified data split. The grid size and spline order are selected based on the best results reported in \autoref{tab_BoT_IoT_f1_grid_contrastive}.}
\label{tab_bot_ablation}
\centering
\begin{tabular}{@{}llll@{}}
\toprule
\textbf{Model} & \textbf{Architecture}       & \textbf{\begin{tabular}{l} Num. of \\ Params \end{tabular}} & \textbf{F1 (\%)} \\ \midrule
KAN (end-to-end)   & {[}20 100 5{]}         & 137,500              & 90.01         \\
KAN (end-to-end)   & {[}20 150 5{]}         & 206,250              & 89.27         \\
KAN (end-to-end)   & {[}20 200 5{]}         & 275,000            & 89.51         \\ \midrule
\textbf{KAN (contrastive)}  & \textbf{{[}20 100 5{]}}         & \textbf{137,500}              & \textbf{93.11}\\
KAN (contrastive)  & {[}20 150 5{]}         & 206,250              & 92.41         \\
KAN (contrastive)  & {[}20 200 5{]}         & 275,000            & 91.56         \\ \midrule
MLP (end-to-end)   & {[}20 220 600 5{]}     & 140,225              & 86.77         \\
MLP (end-to-end)   & {[}20 330 600 5{]}     & 208,535              & 87.23        \\
MLP (end-to-end)   & {[}20 440 600 5{]}     & 276,845            & 87.65         \\ \midrule
MLP (contrastive)  & {[}20 220 600 5{]}     & 140,225              & 89.89         \\ 
MLP (contrastive)  & {[}20 330 600 5{]}    & 208,535              & 88.61         \\ 
MLP (contrastive)  & {[}20 440 600 5{]}     & 276,845            & 89.41         \\ \bottomrule
\end{tabular}
\end{table}

\begin{table}[] 
\caption{Ablation study on the Gas Pipeline dataset using the specified data split. The grid size and spline order are chosen based on the best results in \autoref{tab_gas_pipeline_f1_grid_contrastive}.}
\label{tab_gas_pipeline_ablation}
\centering
\begin{tabular}{@{}llll@{}}
\toprule
\textbf{Model} & \textbf{Architecture}   & \textbf{\begin{tabular}{l} Num. of \\ Params \end{tabular}} & \textbf{F1 (\%)} \\ \midrule
KAN (end-to-end)   & {[}19 100 8{]}    & 21,600                 & 83.23         \\
KAN (end-to-end)   & {[}19 150 8{]}    & 32,400                 & 84.14         \\
KAN (end-to-end)   & {[}19 200 8{]}    & 43,200                 & 84.25         \\ \midrule
\textbf{KAN (contrastive)}  & \textbf{{[}19 100 8{]}}    & \textbf{21,600}                 & \textbf{86.70}\\
KAN (contrastive)  & {[}19 150 8{]}    & 32,400                 & 85.96         \\
KAN (contrastive)  & {[}19 200 8{]}    & 43,200                 & 86.57         \\ \midrule
MLP (end-to-end)   & {[}19 100 180 8{]}& 21,628                 & 77.57         \\
MLP (end-to-end)   & {[}19 155 180 8{]}& 32,628                 & 78.92         \\
MLP (end-to-end)   & {[}19 150 250 8{]}& 42,758                 & 79.12         \\ \midrule
MLP (contrastive)  & {[}19 100 180 8{]}& 21,628                 & 80.13         \\ 
MLP (contrastive)  & {[}19 155 180 8{]}& 32,628                 & 81.78         \\ 
MLP (contrastive)  & {[}19 150 250 8{]}& 42,758                 & 82.21         \\ \bottomrule
\end{tabular}
\end{table}

\subsection{Ablation Study}\label{sub_sec_ablation}
For the ablation study, we focused on two key KAN hyperparameters: spline order and grid size. Although KAN includes several hyperparameters, spline order and grid size are the most influential. The results shown in \autoref{tab_UNSW_15B_f1_grid_contrastive} and \autoref{tab_BoT_IoT_f1_grid_contrastive} indicate that networks with higher grid sizes and smaller spline orders outperform those with higher spline orders. This observation aligns with the findings in \cite{dong2024kolmogorov}, which suggest that layers with larger grid sizes tend to be more robust, despite exhibiting a higher Lipschitz constant. However, this trend does not hold for the Gas Pipeline dataset, as shown in \autoref{tab_gas_pipeline_f1_grid_contrastive}.

Similarly, we examined the performance of KAN against MLP under two training paradigms: contrastive pre-training and end-to-end training. The experiments conducted are summarized in \autoref{tab_unsw_ablation} through \autoref{tab_gas_pipeline_ablation}. For each dataset, we performed 12 experiments to evaluate the effects of the training paradigm and network width. In these tables, two-layer KANs and three-layer MLP networks are compared. For the KAN models, the spline order and grid size parameters were selected based on the best results from \autoref{tab_UNSW_15B_f1_grid_contrastive} to \autoref{tab_gas_pipeline_f1_grid_contrastive}.

In the architecture columns, the first and last numbers denote the input dimensionality and the number of output classes, respectively. For MLPs, the middle number indicates the dimensionality of the hidden layer. Next to the architecture, the total number of parameters for each network is reported. For MLPs, the architectures were chosen to maintain a similar parameter count to their KAN counterparts.

All models, whether trained with contrastive pre-training or end-to-end, were trained using the specified data splits. The main difference is that in the end-to-end training paradigm, only the fine-tuning dataset was used.

It was observed that KAN's performance does not necessarily improve proportionally with the number of parameters. In contrast, MLP networks tend to perform better as the number of parameters increases. This suggests that tuning KAN's hyperparameters is more challenging compared to tuning the parameters of MLP networks. Despite the Gas Pipeline dataset being the most challenging, the parameter counts of models trained on it are significantly lower than those trained on UNSW-NB15 and BoT-IoT.

It is observable that the KAN outperforms the MLP networks, which are tailored to have the same number of parameters as the KAN. In the end-to-end training paradigm, the networks are trained using only the fine-tuning set, without pre-training on unlabeled data. The contrastive training, by leveraging large volumes of unlabeled data, further improves the performance.

It is worth noting that, for KANs, the width of the network has little impact on the performance. As can be observed, the networks with widths of 50 or 100 outperform those with widths of 150 and 200. This supports the earlier findings that the main factors influencing the performance of KAN are the grid size and spline order. However, for MLP networks, increasing the width often results in marginally better performance. This indicates that, for MLP networks, there is a more predictable relationship between network capacity and performance; in other words, MLP networks are easier to tune.

\begin{table*}[]
\caption{Comparison of our proposed methods with other approaches on the UNSW‐NB15 dataset.}
\label{tab_unsw_comparison}
\centering
\begin{tabular}{llllll}
\toprule
\textbf{Literature} & \textbf{Method} & \textbf{Accuracy (\%)} & \textbf{Precision (\%)} & \textbf{Recall (\%)}  & \textbf{F1 (\%)} \\ \midrule
\cite{moustafa2019outlier}                &  Artificial Immune System
                          & 85.00           & 89.00               & 86.00          & 87.00             \\

\cite{moustafa2019outlier}                &  Euclidean Distance Map
                          & 90.00           & 92.00               & 91.00          & 91.00             \\

\cite{moustafa2019outlier}                &  Filter-based SVM
                          & 92.00           & \textbf{94.00}               & 92.00          & 93.00             \\

\cite{al2022new}                          &  Decision Tree
                          & 97.08           &  -                  & -              & 93.21             \\

\cite{al2022new}                          &  Logistic Regression + RQA
                          & 95.75           &  -                  & -              & 90.02             \\ \midrule
Proposed Method                            & Contrastive-KAN                 
                          & \textbf{98.69}             & 89.43          & \textbf{95.85}  & \textbf{93.66}  \\ \bottomrule
\end{tabular}
\end{table*}

\begin{table*}[]
\caption{Comparison of our proposed methods with other approaches on the BoT-IoT dataset.}
\label{tab_bot_iot_comparison}
\centering
\begin{tabular}{llllll}
\toprule
\textbf{Literature} & \textbf{Method} & \textbf{Accuracy (\%)} & \textbf{Precision (\%)} & \textbf{Recall (\%)}  & \textbf{F1 (\%)} \\ \midrule
\cite{van2024two}                &  Deep Neural Network
                          & \textbf{97.57}                      & -               & -          & 93.00             \\
\cite{van2024two}                &  Support Vector Machine
                          & 83.86                      & -               & -          & 42.00             \\
\cite{van2024two}                &  Random Forest
                          & 92.21                      & -               & -          & 56.00             \\
\cite{ferrag2020deep}            &  Convolutional Neural Network   
                          & 97,01                      & -               & -          & -                 \\
\cite{ferrag2021federated}       &  Recurrent Neural Network   
                          & -                          & \textbf{98.20 } & 85.40      & 88.80             \\ \midrule
Proposed Method & Contrastive-KAN                  
                          & 97.52             & 90.41           & \textbf{94.72}  & \textbf{93.11}  \\ \bottomrule
\end{tabular}
\end{table*}

\begin{table*}[]
\caption{Comparison of our proposed methods with other approaches, Gas Pipeline dataset.}
\label{tab_gas_pipeline_comparison}
\centering
\begin{tabular}{llllll}
\toprule
\textbf{Literature} & \textbf{Method} & \textbf{Accuracy (\%)} & \textbf{Precision (\%)} & \textbf{Recall (\%)}  & \textbf{F1 (\%)} \\ \midrule
\cite{feng2017multi}                & Long-Short Term Memory
                          & 87.00               & \textbf{97.00} & 59.00           & 73.00             \\
\cite{feng2017multi}                  & Packet Signature + LSTM 
                          & 92.00                & 94.00         & 78.00           & 85.00             \\
\cite{anthi2021three}                               & Bayesian network
                          & -                    & 84.40         & 85.10           & 84.60              \\  \midrule
Proposed Method & Contrastive-KAN                 
                          & \textbf{91.99}      & 80.16          & \textbf{96.85}  & \textbf{86.70}    \\ \bottomrule
\end{tabular}
\end{table*}
\subsection{Comparative Analysis}
\autoref{tab_unsw_comparison}, \autoref{tab_bot_iot_comparison}, and \autoref{tab_gas_pipeline_comparison} present a comparative analysis of state-of-the-art methods. The performance of the proposed method is superior to other learning-based algorithms, some of which utilize more labeled data than our approach. This demonstrates that the proposed algorithm is well-suited for scenarios with label scarcity.

\section{Conclusion} \label{sec_Conclusion}
In this paper, we proposed a semi-supervised contrastive learning approach for real-time intrusion detection under limited labeled data conditions. This framework demonstrates strong performance on IoT and IIoT datasets, outperforming other state-of-the-art methods. The results of this study show that the performance of the KAN is superior to that of MLP networks when labeled data is scarce. Additionally, contrastive training improves the network's performance by utilizing the available unlabeled data.

A key advantage of the proposed method is the interpretability provided by KAN, in the form of spline function visualization. This feature can be utilized to extract relevant rules from the network, which is beneficial in safety-critical applications, further increasing reliability, stability, and ensuring safe operation.

One of the findings of this paper is that the spline order and grid size are the most influential KAN hyper-parameters to tune, while the layer dimensions have minimal impact on improving performance. For our datasets, a higher grid size and lower spline order yield better results. Another key finding is that hyper-parameter tuning for MLP networks is much easier compared to their KAN counterparts, since KAN does not exhibit a straightforward relationship between performance and the number of parameters or the increase in spline order and grid size.

Although this study performed well on intrusion detection problems with limited labeled data, further studies must be conducted on the KAN architecture for the unsupervised case, where training requires no labeled data. This is left for future work.

\bibliographystyle{unsrt}  
\bibliography{references}

\end{document}